\documentclass[11pt]{article}
\usepackage{epsf,amsmath,amssymb,graphicx,scalefnt,subfigure,ulem}
\usepackage{graphicx}
\usepackage{psfrag}  
\usepackage{amsfonts}
\usepackage{hyperref}
\usepackage{cite,setspace}
\usepackage[small,bf]{caption}

\newcommand{\pdf}{{\abbrev PDF}}
\newcommand{\qcd}{{\abbrev QCD}}
\newcommand{\abbrev}{\scalefont{.9}}
\newcommand{\muF}{\mu_{\rm F}}
\newcommand{\muR}{\mu_{\rm R}}
\newcommand{\mH}{m_{\rm H}}
\newcommand{\ep}{\epsilon}
\newcommand{\lhc}{{\abbrev LHC}}
\newcommand{\sm}{{\abbrev SM}}
\newcommand{\mssm}{{\abbrev MSSM}}
\newcommand{\susy}{{\abbrev SUSY}}

\newcommand{\lo}{{\abbrev LO}}
\newcommand{\nlo}{{\abbrev NLO}}

\newcommand{\nnlo}{{\abbrev NNLO}}

\def\be{\begin{equation}}
\def\ee{\end{equation}}
\def\bea{\begin{eqnarray}}
\def\eea{\end{eqnarray}}
\def\benn{\begin{equation*}}
\def\eenn{\end{equation*}}
\def\beann{\begin{eqnarray*}}
\def\eeann{\end{eqnarray*}}
\def\ep{\epsilon}

\def\nn{\nonumber}
\def\d{\mathrm{d}}
\def\Li2{{\rm Li}_2}

\def\fb{f_b(\bar{x}_1^0)}
\def\fbb{f_{\bar{b}}(\bar{x}_2^0)}

\graphicspath{{./diagrams/}}

\begin{document}

\begin{titlepage}

\begin{flushright} 
          UCLA/10/TEP/108\\
          WUB/10-25
\end{flushright}

\begin{center}
\begin{Large}
{\bf Analytic Results for Higgs Production in Bottom Fusion} 
\end{Large}

\vskip 1.5cm

Kemal J. Ozeren\\

\vskip 0.7cm

{\it Department of Physics and Astronomy} \\
{\it UCLA, Los Angeles, CA 90095-1547} \\
{\small {\tt ozeren@physics.ucla.edu}} 

\vskip 0.4cm
and 
\vskip 0.4cm

{\it Fachbereich C, Bergische Universit\"at Wuppertal} \\
{\it 42097 Wuppertal, Germany} \\

\end{center}

\vskip 2 cm 

\begin{abstract}
We evaluate analytically the cross section for Higgs production plus one jet through bottom quark fusion. By considering the small $p_T$ limit we derive expressions for the resummation coefficients governing the structure of large logarithms, and compare these expressions with those available in the literature.
\end{abstract}

\end{titlepage}

\section{Introduction}

The Standard Model (\sm{}) \cite{Glashow:1961tr,Weinberg:1967tq} predicts the existence of a massive scalar particle known as the Higgs boson. Within this theory and its supersymmetric extensions \cite{Nilles:1983ge}, matter fields and gauge bosons acquire mass via the Higgs mechanism \cite{Guralnik:1964eu,Englert:1964et,Higgs:1964pj}. Although the Higgs remains undiscovered, experiments have placed restrictions on its mass \cite{Barate:2003sz,Aaltonen:2010yv}. Supersymmetric theories require more than one Higgs boson. There are many more free parameters than in the \sm, and the experimental constraints are correspondingly weaker \cite{Schael:2006cr,Benjamin:2010xb}.

\begin{figure}[t]
\centering
\subfigure[]{\label{fig:fourFS}}{\includegraphics[width=6em]{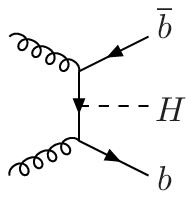}}
\hspace{1.5in}
\subfigure[]{\label{fig:fiveFS}}{\includegraphics[width=7.5em]{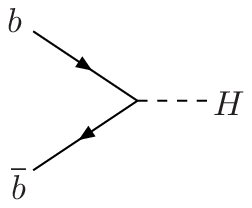}}
\caption{Diagrams for Higgs production through bottom fusion in the (a) four and (b) five flavour schemes.}
\end{figure}

The Large Hadron Collider (\lhc{}) is expected to find the Higgs boson if it exists. To achieve this, various production mechanisms must be considered. The relative utility of each depends strongly on the Higgs' mass and couplings. While in the \sm{} gluon fusion is by far the largest contribution to the total cross section, in \susy{} theories with large $\tan \beta$ bottom quark fusion can dominate, due to the enhanced $b\bar{b}H$ Yukawa coupling \cite{Belyaev:2005ct,Belyaev:2005nu,Brein:2010xj}. Reviews can be found in Refs.~\cite{Djouadi:2005gi,Djouadi:2005gj}. If we assume that the proton is composed only of the four lightest quarks and the gluon, the so called four flavour scheme ({\abbrev 4FS}), then the dominant leading order diagram for this process is that shown in Fig.~\ref{fig:fourFS}. Integration of the phase space leads to divergences arising from the kinematical region where one or both bottom quarks are collinear to the initial state partons. The bottom's mass $m_b$ regulates these divergences, but they still leave traces in the form of large logarithms $\ln(m_b^2/\mH^2)$. Such logarithms jeopordise the convergence of the perturbative series, so ideally one would like to resum them. This can be achieved by introducing bottom quark \pdf s. In this five flavour scheme ({\abbrev 5FS}) \cite{Barnett:1987jw,Dicus:1988cx} the $b$ quark can appear in the initial state, and so the leading order process is changed to that appearing in Fig.~\ref{fig:fiveFS}. One sets the $b$ quark mass to zero in this case. Results obtained in either scheme should be the same, although when truncating at a finite order there will be differences, formally of higher order in $\alpha_s$. Despite this, it was found that the inclusive cross sections in the {\abbrev 4FS} and the {\abbrev 5FS} differ by roughly a factor of five when evaluated at $\muF=\muR=\mH$, where $\muF/\muR$ is the factorisation/renormalisation scale.  This remains true also at \nlo{} \qcd{} which was calculated for the {\abbrev 5FS} in Refs.\,\cite{Dicus:1998hs,Maltoni:2003pn}, and for the {\abbrev 4FS} in Refs.\,\cite{Dittmaier:2003ej,Dawson:2003kb}.  It was thus proposed in Refs.\,\cite{Rainwater:2002hm,Plehn:2002vy,Maltoni:2003pn,Boos:2003yi} that when using the five flavour scheme the appropriate central scale is $m_H/4$. Indeed, the \nnlo{} result \cite{Harlander:2003ai} in the {\abbrev 5FS} seems to confirm this choice.

Higgs production in association with one or more jets \cite{Ellis:1987xu} has also received much attention. In the case of gluon fusion the leading order cross section is known, including the full top and bottom mass dependence, in both the \sm{} \cite{Ellis:1987xu,Field:2003yy,Brein:2003df} and \mssm{} \cite{Brein:2003df}. The \nlo{} corrections are known only in the heavy-top limit \cite{deFlorian:1999zd,Ravindran:2002dc,Glosser:2002gm}. As far as the \mssm{} is concerned, one expects that to a very good approximation one can simply replace the effective $ggH$ coupling of the Standard Model with its \mssm{} value \cite{Harlander:2003bb,Harlander:2004tp,Degrassi:2008zj}. However, as we have stressed, for large $\tan \beta$ one must also include the bottom fusion contribution. That is the subject of this paper.

The bottom fusion contribution to $H+$jet production has been considered for the case in which a final state $b$ quark is tagged \cite{Campbell:2002zm}. This is a useful observable because one can measure the $b\bar{b}H$ Yukawa coupling directly. Without $b$ tagging, this process is a contribution to the total $H+$jet cross section, and must be considered alongside gluon fusion. For this case various distributions at \nlo{} have been presented \cite{Harlander:2010cz}, based on Catani-Seymour subtraction. In this paper we give the same cross section analytically. As well as providing a very strong check on the results of Ref.~\cite{Harlander:2010cz}, our results allow us to analytically take the $p_T \to 0$ limit, and thus derive expressions for the resummation coefficients which govern the structure of large logarithms. This will be described in Section \ref{smallpt}.

\section{Notation}

We are interested in the transverse momentum distribution of Higgs bosons arising from the scattering of two hadrons $h_1$ and $h_2$ at centre of mass energy $\sqrt{S}$. In particular, we consider in this paper only that part of the cross section proportional to the $Hb\bar{b}$ Yukawa coupling. The $b$-quark mass is set to zero everywhere except in this coupling. For a Higgs boson with tranverse momentum $p_T$ and rapidity $y$, the cross section is
\be \label{xsec}
\frac{\d\sigma}{\d p_T^2\d y} = \sum_{ij} \frac{1}{1+\delta_{ij}}\int \d x_1 \int \d x_2 f_{i/h_1}(x_1,\muF) f_{j/h_2}(x_2,\muF) \frac{\d\hat{\sigma}_{ij}}{\d p_T^2\d y},
\ee
where $f_{i/h_1}(x_1,\muF)$ is the parton density for finding a parton $i$ in hadron $h_1$. We expand the partonic cross section appearing on the right hand side in powers of the strong coupling constant $\alpha_s(\muR)$,
\be
\frac{\d\hat{\sigma}_{ij}}{\d p_T^2\d y} = \frac{\pi}{8} \frac{m_b^2}{\mathcal{V}^2} \frac{1}{\hat{s}} \frac{1}{\mathcal{C}_{ij}} \left[\frac{\alpha_s(\muR)}{2\pi}G^{(1)}_{ij}(\mu_R) + \left(\frac{\alpha_s(\muR)}{2\pi}\right)^2 G^{(2)}_{ij}(\mu_R)  + \cdots \right],
\ee
where $\mathcal{V} = 246$ GeV is the vacuum expectation value of the Higgs field, $m_b$ is the bottom quark mass, $\mathcal{C}_{ij}$ is a colour averaging factor ($\mathcal{C}_{ij} = 9$ for quark-quark scattering, etc.) and the dots stand for higher terms in the $\alpha_s$ expansion. Although the partonic cross section itself is not a function of the renormalisation scale $\muR$, its expansion coefficients $G^{(n)}_{ij}$ are, so that truncating at any finite order of perturbation theory leads to an unphysical $\muR$ dependence of the cross section. Reducing this unphysical scale dependence is one of the primary motivations for calculating higher order \qcd{} corrections.

We denote the four momenta of the incoming hadrons as $P_1$ and $P_2$, while those of the colliding partons are $p_1=x_1P_1$ and $p_2=x_2P_2$. The mass, transverse momentum and rapidity of the Higgs boson are written $\mH$, $p_T$ and $y$ respectively. Momentum conservation implies
\be
p_1 + p_2 = Q + p_H,
\ee
where $Q$ represents the total momentum of the final state \qcd{} partons, of which there can be either one or two. These we will label $p_3$ and $p_4$. Our results for the coefficients $G^{(n)}_{ij}$ will be given in terms of the following partonic invariants
\bea
\nn s &=& (p_1+p_2)^2, \\
u &=& (p_1-Q)^2, \\
\nn t &=& (p_2-Q)^2,
\eea
in terms of which momentum conservation imposes the constraint
\be
s + u + t = \mH^2 + Q^2.
\ee
In terms of these variables the transverse momentum of the Higgs satisfies
\be
p_T^2 = \frac{ut-\mH^2Q^2}{s}.
\ee
It is also useful to define 
\bea
\nn S_u &=& u-Q^2, \\
\label{SuStv} S_t &=& t-Q^2, \\
\nn m_T^2 &=& \mH^2 + p_T^2, \\
\nn v &=& \frac{p_T^2}{Q^2+p_T^2}.
\eea
At leading order ($\mathcal{O}(\alpha_s)$) there are two contributing channels. We find for the corresponding coefficients $G^{(1)}_{ij} = g_{ij} \delta(Q^2)$ with,
\begin{align}
\label{gbb} g_{b\bar{b}} &= 4\,C_FC_A~\frac{(s^2+\mH^4)}{ut}, \\
g_{bg} &= 4\,C_FC_A~\frac{(u^2+\mH^4)}{-st},
\end{align}
where $C_F=\frac{4}{3}$ and $C_A=3$. We will discuss how to integrate these expressions over the momentum fractions $x_{1,2}$ in section \ref{results}.

\section{{\abbrev NLO} Corrections}
\begin{figure}[htp]
\centering
\subfigure{\includegraphics{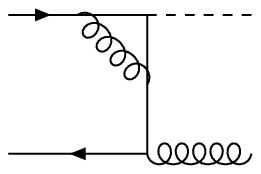}}
\hspace{0.5in}
\subfigure{\includegraphics{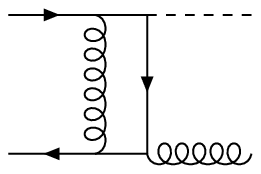}}
\caption{Examples of one loop diagrams at \nlo{}}
\label{fig:virtual}
\end{figure}
The $\mathcal{O}(\alpha_s^2)$ corrections, which form the coefficient function $G^{(2)}_{ij}$, receive three different types of contribution: the one loop virtual corrections to the leading order processes, for which some sample diagrams are given in Fig.~\ref{fig:virtual}, the mass factorisation pieces, arising from the definition of the parton densities at \nlo{}, and finally the real radiation contribution. Each of these pieces is divergent in four dimensions, so that in practice a regularisation procedure is required. We use conventional dimensional regularisation (CDR), working in $d=4-2\ep$ dimensions, so that the divergences manifest themselves as poles in the parameter $\ep$. For infrared safe observables these poles cancel, and at the end of the calculation we may safely take the limit $d \to 4$.

The one loop amplitude for $b\bar{b} \to Hg$ was given in Ref.~\cite{Campbell:2002zm}, and we have independently checked the result. We also require $bg \to Hb$, which can be obtained by crossing. The virtual parts contain ultraviolet divergences, which we remove by renormalising $\alpha_s$ and $m_b$ in the $\overline{\mbox{MS}}$ scheme. 

At \nlo{} there are also additional channels to consider, beyond those which contribute at \lo{}. In our case these are $gg$, $bq$, $bb$ and $q\bar{q}$. Here $q$ represents one of the $u$, $d$, $s$ or $c$ quarks, and it is understood that the charge conjugated processes are also included. Because these channels do not contribute at leading order, none of them have a one loop correction. However, with the exception of $q\bar{q}$, they do require mass factorisation to remove collinear poles, and so must be regularised just as in the case of $b\bar{b}$ and $bg$. 

\subsection{Real Radiation}

\begin{figure}[htp]
\centering
\subfigure{\includegraphics{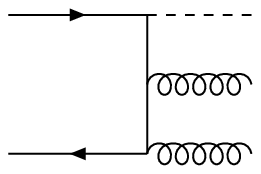}}
\hspace{0.5in}
\subfigure{\includegraphics{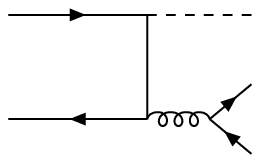}}
\caption{Examples of real emission diagrams at \nlo{}. The initial state lines are bottom quarks, but the final state quarks can be of any flavour.}
\label{fig:real}
\end{figure}

We have evaluated the amplitudes for the Higgs plus two parton processes using FORM \cite{Vermaseren:2000nd}. Sample diagrams are shown in Fig.~\ref{fig:real}. They are expressed in terms of the invariants $s_{ij} = (p_i+p_j)^2$ and $s_{ijk} = (p_i+p_j+p_k)^2$. Note that we must retain the $\mathcal{O}(\ep)$ pieces of the amplitudes, because integration over the three body phase space can generate poles in $\ep$. As a check one may verify that gauge invariance holds. We have also used {\tt MADGRAPH} \cite{Stelzer:1994ta} to check the amplitudes in the $\ep \to 0$ limit.

We write the three body phase space factor as
\be \label{PS3}
\d\Gamma_3 = \left( \frac{4\pi \mu^2}{Q^2} \right)^{\ep} \left( \frac{4\pi \mu^2}{p_T^2} \right)^{\ep} \frac{1}{(4\pi)^2} \, \frac{1}{\Gamma(1-2\ep)} \, \frac{\d\Omega}{8\pi} \, \d p_T^2 \, \d y,
\ee
and integrate analytically over the angular factor $\d\Omega$, given by
\be
\int \d\Omega = \frac{1}{2\pi} \int^{\pi}_0 \sin^{1-2\ep}\theta~\d\theta \int^{\pi}_0 \sin^{-2\ep}\phi~\d\phi.
\ee
It is useful to work in the $Q$ rest frame. We can then obtain expressions \cite{Ravindran:2002dc} for the invariants in terms of angles and energies. To perform the integration we first make use of momentum conservation and numerous partial fraction identities to ensure that each term in the amplitude squared contains at most two invariants that depend on the angles $\theta$ and $\phi$. These invariants are $s_{13}$, $s_{14}$, $s_{23}$, $s_{24}$, $s_{123}$ and $s_{124}$. For example, the relation
\be
\frac{1}{s_{13} \, s_{14}} = \frac{1}{S_u} \left( \frac{1}{s_{13}} + \frac{1}{s_{14}}\right),
\ee
reduces the number of angle dependent factors from two to one. Once this decomposition is achieved the integrals over $\d\Omega$ are of two types ($m$ and $n$ are integers):
\bea
\int d\Omega \ s_{23}^{-m} \, s_{13}^{-n}, \\
\int d\Omega \ s_{123}^{-m} \, s_{13}^{-n}.
\eea
A closed form result, valid for abritrary $\ep$, for integrals of the first type is given in Ref.~\cite{vanNeerven:1985xr}. The integrals of the second type are given as expansions in $\ep$ in Ref.~\cite{Beenakker:1988bq}. We have used these results directly, supplementing them where necessary with extra $\mathcal{O}(\ep)$ terms.

As the integration over the momentum fractions $x_{1,2}$ is performed, divergences appear in the small $Q^2$ region due to terms with $1/Q^2$ factors. These divergences are regulated\footnote{Strictly speaking, we mean `made integrable'.} by the $Q^{-2\ep}$ factor appearing Eq.\eqref{PS3}. To expose the corresponding poles in $\ep$ we use the distribution relation
\be \label{Q2distid}
(Q^2)^{-1-\ep} \to -\frac{1}{\ep} \delta(Q^2) A^{\ep} + \left(\frac{1}{Q^2}\right)_+ - \ep \left(\frac{\ln Q^2}{Q^2}\right)_+ + \mathcal{O}(\ep^2),
\ee
where $A$ is the maximum value of $Q^2$. The plus distributions above will appear in our final results. They are defined by
\be
\int^A_0 \d Q^2 \, f(Q^2) \left[ g(Q^2) \right]_+ = \int_0^A \d Q^2 \, \left[f(Q^2) - f(0)\right] g(Q^2).
\ee
There are still divergences in the small $p_T$ region. These are regulated by the $p_T^{-2\ep}$ factor appearing in Eq.\eqref{PS3}. In principle we could make use of a distribution relation similar to that above to expose the divergences as poles in $\ep$. We would then need to add the two loop corrections to the process $b\bar{b} \to H$, as well as some extra terms arising from mass factorisation. The result would be a \nnlo{} result for the differential cross section of the process $b\bar{b} \to H$. We do not take this extra step, which is difficult to achieve in practice, so that in our numerical results we must avoid the small $p_T$ region. Our results therefore constitute a \nlo{} result for the process $b\bar{b} \to H + \textrm{jet}$. We will discuss the behaviour of the cross section in the $p_T \to 0$ limit in more detail in Section \ref{smallpt}.

\section{Results} \label{results}

Our result for the \nlo{} $b\bar{b}$ coefficient function takes the form
\be
G^{(2)}_{b\bar{b}} = D_{b\bar{b}} \delta(Q^2) + E_{b\bar{b}} \left(\frac{1}{Q^2}\right)_+ + F_{b\bar{b}} \left(\frac{1}{Q^2} \ln \frac{Q^2}{\mH^2}\right)_+ + H_{b\bar{b}}.
\ee
We define 
\bea
\kappa &=& \frac{1}{2}\left(\mH^2+s-Q^2\right), \\
\lambda^2 &=& \kappa^2 - \mH^2 s, \\
x &=& \frac{\kappa+\lambda}{\kappa-\lambda},
\eea
and introduce the following convenient logarithm abbreviations.
\bea
\nn L_s &=& \ln \frac{s}{\mH^2}, \qquad L_u = \ln \frac{-u}{\mH^2}, \qquad L_t = \ln \frac{-t}{\mH^2},\\
l_F &=& \ln \frac{\muF^2}{\mH^2}, \qquad l_R = \ln \frac{\muR^2}{\mH^2}, \qquad L_A = \ln \frac{A}{\mH^2}.
\eea
We give here the $D_{b\bar{b}}$, $E_{b\bar{b}}$ and $F_{b\bar{b}}$ coefficients. The expressions for $H_{b\bar{b}}$ are very large, so we do not reproduce them here. Results for all the above coefficients, for all channels, are available from the author on request.
\begin{align}
\nn D_{b\bar{b}} &=  g_{b\bar{b}} C_A \bigg[  - L_A^2 - 2 L_A L_s + 2 L_A L_t + 2 L_A L_u - \frac{11 L_A}{6} + \frac{11 l_R}{6} - L_s^2 
+2 L_s L_t \\
\nn  & + 2 L_s L_u - L_t^2 - 2 L_t L_u - L_u^2 -2 \Li2\left(1-\frac{\mH^2}{s}\right)+\frac{67}{18} \bigg] \\
\nn  &+ g_{b\bar{b}} C_F \bigg[ 4 L_A^2 + 4L_A L_s - 4L_A L_t - 4L_AL_u - 4 L_A l_F + 3l_R + 2L_s^2 \\ 
\label{Dres} &- 4L_s L_t -4L_s L_u + L_t^2 + 2L_t L_u +2 L_tl_F + L_u^2 + 2L_ul_F -3l_F \\ 
\nn  & + 4\Li2\left(1-\frac{\mH^2}{s}\right) + 2\Li2\left(\frac{\mH^2}{\mH^2-t}\right) + 2\Li2\left(\frac{\mH^2}{\mH^2-u}\right) \\
\nn  &+ \ln^2\left(1-\frac{t}{\mH^2}\right) + \ln^2\left(1-\frac{u}{\mH^2}\right) + \frac{\pi^2}{3} - 2 \bigg] \\
\nn  &+ \frac{2}{3} n_f T_R g_{b\bar{b}} \bigg[L_A - l_R - \frac{5}{3} \bigg] + 4\mH^2\left(\frac{1}{u}+\frac{1}{t}\right)  C_FC_A(C_A - C_F), \\
E_{b\bar{b}} &= C_A g_{b\bar{b}} \bigg[ \ln(1-v) + \ln\frac{S_u}{u}+ \ln\frac{S_t}{t} + 2\ln \frac{Q^2+p_T^2}{\mH^2} - L_s - \frac{11}{6}) \bigg] \\
\nn &+ C_Fg_{b\bar{b}} \bigg[-2\ln\frac{S_u}{u} -2 \ln\frac{S_t}{t} - 4 l_F -4\ln \frac{Q^2+p_T^2}{\mH^2} + 2 L_s \bigg] + \frac{2}{3}T_R n_F g_{b\bar{b}} \\
\nn & + 4 C_F C_A\left(C_F-\frac{C_A}{2}\right) \frac{\ln x}{\lambda} \bigg[ 2\frac{\mH^4}{t} + \frac{\mH^4}{u} -2\frac{\mH^2 u}{t} -\mH^2 + \frac{s^2}{u} - s + \frac{u^2}{t} + u \bigg], \\
F_{b\bar{b}} &= 8\left(C_F - \frac{C_A}{4}\right) g_{b\bar{b}}.
\end{align}
Recall that $S_u$, $S_t$ and $v$ were defined in Eq.~\eqref{SuStv}, while $g_{b\bar{b}}$ was given in Eq.~\eqref{gbb}. $\Li2$ denotes the dilogarithm function. For \qcd{} the colour factors take the values $C_F = \frac{4}{3}$, $C_A = 3$ and $T_R = \frac{1}{2}$. 

To integrate the distributions above over the \pdf s, it is useful to arrange for $Q^2$ to be one of the integration variables. This is easily achieved \cite{Glosser:2001nv}, with the result that we can replace
\be
\int_0^1 \d x_1 \int_0^1 \d x_2 ~ \theta(Q^2) \to \frac{1}{S} \int^1_{x_+} \frac{\d x_1}{x_1 - x_U} \int_0^{A_1} \d Q^2 + \frac{1}{S} \int^1_{x_-} \frac{\d x_2}{x_2 - x_T} \int_0^{A_2} \d Q^2,
\ee
where
\beann
x_U &=& \frac{m_T^2}{S} e^{y}, \\
x_T &=& \frac{m_T^2}{S} e^{-y}, \\
x_{\pm} &=& \frac{m_T+p_T}{\sqrt{S}} e^{\pm y}, \\
A_1 &=& x_1(1-x_T) - x_U + \frac{\mH^2}{S}, \\
A_2 &=& x_+(x_2-x_T) - x_U x_2 + \frac{\mH^2}{S}. \\
\eeann
The expression for $A_1$ or $A_2$, as appropriate, should be substituted in Eq.~\eqref{Dres} in place of $A$. In this way one can obtain numerical results for the hadronic cross section. Phenomenological analyses have already been presented in Ref.~\cite{Harlander:2010cz}, so we do not repeat it here.

\subsection{Checks}

We have performed a number of checks on our results. Firstly, the dependence of the cross section on $\muF$ and $\muR$ can be predicted due to the requirement that physical observables must be independent of these scales. Our expressions for the perturbative coefficients $G^{(n)}_{ij}$ satisfy these constraints. Secondly, as we will discuss in the next section, the small $p_T$ behaviour can be compared to known resummed formulae.

The strongest check is a comparison with a Monte Carlo numerical code \cite{Harlander:2010cz} based on the Catani Seymour \cite{Catani:1996vz} subtraction formalism. We find excellent agreement for all channels. In the case of the $q\bar{q}$ channel, one can also compare to the known total cross section \cite{Harlander:2003ai} by numerically integrating Eq.\eqref{xsec} over $p_T$ and $y$.

\section{Small $p_T$ limit} \label{smallpt}

For small $p_T$ the convergence of fixed order perturbation theory is spoiled at any finite order by terms of the form
\be
\frac{\ln^n p_T^2}{p_T^2}, \quad n=0,1,2,3.
\ee
To obtain reliable physical predictions for observables in this region one must resum these enhanced terms. The method to perform this resummation is known. The result \cite{Collins:1984kg} is expressed as an integral over the impact parameter $b$,
\be \label{resum}
\frac{\d\sigma}{\d p_T^2 \d y} = \frac{\mH^2 \sigma_0(\mH)}{2S} \int^{\infty}_0 b \, \d b \, J_0(b p_T) W(b), 
\ee
where the bottom quark mass $m_b$, implicit in the prefactor $\sigma_0$, is evaluated at the scale $\mu_R = \mH$. The Sudakov form factor $W(b)$ contains the large logarithms, and is defined as
\begin{align}
\nonumber W(b) = &\left(C_{bi}(\alpha_s(b_0/b)) \otimes f_i \right) (\bar{x}_1^0;b/b_0) \left(C_{\bar{b}j}(\alpha_s(b_0/b)) \otimes f_j \right) (\bar{x}_2^0;b/b_0) \\
 & \times \exp \left\{- \int^{\mH^2}_{b_0^2/b^2} \frac{\d q^2}{q^2}\left[A(\alpha_s(q)) \ln \frac{\mH^2}{q^2} + B(\alpha_s(q)) \right]  \right\}
\end{align}
where $\otimes$ indicates convolution, partons $i,j$ are implicitly summed over and $b_0=2e^{-\gamma_E}$, with $\gamma_E$ Euler's constant. The \pdf s $f_i$ and $f_j$ are evaluated at the scale $b/b_0$, and
\be
\bar{x}^0_{1,2} = \frac{\mH}{\sqrt{S}} e^{\pm y}.
\ee
 The resummation coefficients $A$, $B$ and $C_{ij}$ can be expanded perturbatively,
\bea
A(\alpha_s) &=& \sum_{n=1}^{\infty} \left(\frac{\alpha_s}{2\pi}\right)^n A^{(n)}, \\
B(\alpha_s) &=& \sum_{n=1}^{\infty} \left(\frac{\alpha_s}{2\pi}\right)^n B^{(n)}, \\
C_{ij}(\alpha_s) &=& \delta_{ij} \delta(1-z) + \sum_{n=1}^{\infty} \left(\frac{\alpha_s}{2\pi}\right)^n C_{ij}^{(n)}.
\eea
The coefficient $A^{(1)}$ controls the leading logarithmic ({\abbrev LL}) terms, while $A^{(2)}$, $B^{(1)}$ and $C_{ij}^{(1)}$ give the next to leading logarithmic ({\abbrev NLL}) terms, etc. They can be evaluated by performing a fixed order calculation and comparing to the resummed expression.

\subsection{Fixed Order Expansion of the Resummed Formula}

One cannot naively expand Eq.~\eqref{resum} in powers of $\alpha_s$. Instead, we first integrate by parts (we can ignore the surface term) to obtain
\be
\frac{\d\sigma}{\d p_T^2 \d y} = -\frac{\mH^2 \sigma_0(\mH)}{2S} \frac{1}{p_T^2} \int^{\infty}_0 b \,  \d b \, J_1(b) \frac{\d W(b)}{\d b}.
\ee
With the $p_T^2$ pole now manifest, we can expand. We use the {\abbrev DGLAP} equation to evolve the \pdf s to an arbitrary scale $\muF$. We also evolve the \qcd{} coupling $\alpha_s$ and the bottom quark mass $m_b$ from their values at the scale $\mH$ or $q$ to an arbitrary scale $\muR$. The expanded resummed cross section is, using the notation of Ref.~\cite{Glosser:2002gm}, 
\be \label{expres}
\frac{\d\sigma}{\d p_T^2 \d y} \Bigg|_{p_T\ll m} = \frac{\sigma_0}{S}\frac{\mH^2}{p_T^2} \left[ \sum_{m=1}^2 \sum_{n=0}^{2m-1} \left(\frac{\alpha_s}{2\pi}\right)^m \phantom{}_mC_n \left(\ln \frac{\mH^2}{p_T^2} \right)^n + \mathcal{O}(\alpha_s^3) \right].
\ee
The coefficients $_mC_n$ are related to the resummation coefficients as follows,
\beann
_1C_1 &=& A^{(1)} \fb \fbb, \\
_1C_0 &=& B^{(1)} \fb \fbb + (P_{bi} \otimes f_i)(\bar{x}_1^0) \fbb + \fb (P_{\bar{b}i} \otimes f_i)(\bar{x}_2^0), \\
_2C_3 &=& -\frac{1}{2}\left[A^{(1)}\right]^2 \fb \fbb, \\
_2C_2 &=& -\frac{3}{2}A^{(1)} \bigg[(P_{bi} \otimes f_i)(\bar{x}_1^0) \fbb + \fb (P_{\bar{b}i} \otimes f_i)(\bar{x}_2^0)\bigg] \\
&&+ A^{(1)} \left[\beta_0 - \frac{3}{2}B^{(1)}\right] \fb \fbb, \\
_2C_1 &=& \left[\beta_0 - 2 B^{(1)} - A^{(1)} \ln\frac{\muF^2}{\mH^2}\right] (P_{bi} \otimes f_i)(\bar{x}_1^0) \fbb + A^{(1)} (C_{bi}^{(1)} \otimes f_i) (\bar{x}_1^0) \fbb  \\
&& - (P_{bi}\otimes f_i)(\bar{x}_1^0) (P_{\bar{b}j}\otimes f_j)(\bar{x}_2^0) - (P_{bi}\otimes P_{ij} \otimes f_j)(\bar{x}_1^0) \fbb \\
&& - \frac{1}{2} \left[\left[B^{(1)}\right]^2 - A^{(2)} - \beta_0B^{(1)} - \beta'_0 A^{(1)}\ln\frac{\muR^2}{\mH^2}  \right] \fb \fbb \\
&&+ \{b,\bar{x}_1^0 \leftrightarrow \bar{b}, \bar{x}_2^0 \} \\
_2C_0 &=& -\bigg[(P_{bi}\otimes f_i)(\bar{x}_1^0) (P_{\bar{b}j}\otimes f_j)(\bar{x}_2^0) + (P_{bi}\otimes P_{ij} \otimes f_j)(\bar{x}_1^0) \fbb \bigg] \ln\frac{\muF^2}{\mH^2} \\
&& + \bigg[ \beta'_0 \ln\frac{\muR^2}{\mH^2} - B^{(1)}\ln\frac{\muF^2}{\mH^2}  \bigg] (P_{bi} \otimes f_i)(\bar{x}_1^0) \fbb \\
&&+(C_{bi}^{(1)} \otimes f_i)(\bar{x}_1^0) (P_{\bar{b}j} \otimes f_j)(\bar{x}_2^0) + (C_{bi}^{(1)}\otimes P_{ij} \otimes f_j)(\bar{x}_1^0) \fbb \\
&&+\left[\zeta_3 \left[A^{(1)}\right]^2 + \frac{1}{2}B^{(2)} + \frac{1}{2}\beta'_0 B^{(1)}\ln\frac{\muR^2}{\mH^2} \right] \fb \fbb \\
&&+ (B^{(1)} - \beta_0) (C_{bi}^{(1)} \otimes f_i)(\bar{x}_1^0)\fbb + (P^{(2)}_{bi}\otimes f_i)(\bar{x}_1^0) \fbb \\
&&+ \{b, \bar{x}_1^0 \leftrightarrow \bar{b},\bar{x}_2^0 \}.
\eeann
where $\zeta_n$ is the Riemann $\zeta$-function ($\zeta_3 = 1.202 \dots$) and $\beta_0 = (11C_A - 4T_R n_f)/6$. The corresponding expansions for Drell Yan production \cite{Arnold:1990yk} and Higgs production through gluon fusion \cite{Glosser:2002gm} have been presented before. Our case is slightly different to each of these due to the $\muR$ dependence of the $Hb\bar{b}$ Yukawa coupling. This is reflected in the modified beta coefficient $\beta'_0 = \beta_0 + 3C_F$. The two loop splitting function $P^{(2)}_{b\bar{b}}$ can be extracted from the results of \cite{Curci:1980uw,Furmanski:1980cm}.

\subsection{Extracting the Resummation Coefficients}

We have checked analytically that in the limit of small Higgs transverse momentum our results reproduce the resummed result, when the latter is expanded to the appropriate order in $\alpha_s$. Taking the limit analytically requires great care, because as well as explicitly singular terms appearing in our results, some logarithms of $p_T$ appear only upon integration over the momentum fractions $x_{1,2}$.

\begin{figure} 
\centering
\includegraphics{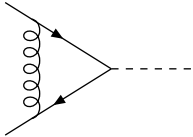}
\caption{Diagram controlling the process-dependent part of the {\abbrev NNLL} resummation coefficient $B^{(2)}$, as described in Ref.~\cite{deFlorian:2001zd}.}
\label{fig:B2diag}
\end{figure}

By comparing with Eq.~\eqref{expres} we can derive the values of the resummation coefficients $A$, $B$ and $C_{ij}$. For the universal coefficients we find the expected values,
\bea
A^{(1)} &=& 2C_F, \\
A^{(2)} &=& 2C_F\left(\frac{67}{18}C_A - \frac{10}{9}n_fT_R - \frac{\pi^2}{6}C_A\right), \\
B^{(1)} &=& -3C_F.
\eea
For the process-specific coefficients we have
\bea
\nn B^{(2)} &=&  C_F^2 \left(-\frac{3}{4} + \pi^2 -12\zeta_3 \right) + C_FC_A\left( -\frac{61}{12} +\frac{11}{9}\pi^2 +6\zeta_3 \right) + C_F T_R n_f \left(\frac{5}{3} - \frac{4}{9}\pi^2 \right), \\
C^{(1)}_{b\bar{b}} &=& C_F\left[1-x + \left(\frac{\pi^2}{2}-1 \right)\delta(1-x)  \right],\\
\nn C^{(1)}_{bg} &=& 2 \, T_R \, x(1-x).
\eea
The expressions for $C^{(1)}_{ij}$ match those given in Ref.~\cite{Belyaev:2005bs}. Our result for $B^{(2)}$ is the first direct calculation of this quantity. It has been shown \cite{deFlorian:2001zd} that $B^{(2)}$ can be split into universal and process dependent parts, and that furthermore, the process dependent part is directly related to the finite part $\mathcal{A}$ of the one loop correction to the leading order process, which in our case is $b\bar{b}\to H$. For quark initiated processes, the relationship is expressed as
\be \label{grazziniB2}
B^{(2)} = -2\gamma^{(2)} + \beta_0 \left( \frac{2}{3}C_F \pi^2 + \mathcal{A} \right),
\ee
where $\gamma^{(2)}$ is the coefficient of $\delta(1-z)$ in the two loop splitting function $P^{(2)}_{q\bar{q}}(z)$, given by
\be
\gamma^{(2)} = C_F^2\left(\frac{3}{8}-\frac{\pi^2}{2}+6\zeta_3 \right) + C_FC_A\left(\frac{17}{24} + \frac{11}{18}\pi^2 -3\zeta_3\right) - C_F n_F T_R \left(\frac{1}{6}+\frac{2}{9}\pi^2 \right).
\ee 
It is straightforward to evaluate the one loop correction to $b\bar{b} \to H$ (the single contributing diagram is shown in Fig.~\ref{fig:B2diag}), from which we find
\be
\mathcal{A} = C_F\left( -2 + \frac{2}{3}\pi^2 \right).
\ee

Substituting this into Eq.~\eqref{grazziniB2} yields the same expression for $B^{(2)}$ as we have derived from our analytic form of the cross section.

\section{Conclusion}

We have described the analytic calculation of the cross section $b\bar{b} \to H + \textrm{jet}$. The partonic cross section is a distribution in $Q^2$, the invariant mass of the final state \qcd{} partons. The results agree numerically with an implementation based on Catani-Seymour subtraction \cite{Harlander:2010cz}.

By taking the limit of small Higgs transverse momentum, we have evaluated the resummation coefficients that govern the structure of large logarithms, including the {\abbrev NNLL} coefficient $B^{(2)}$. This is the first direct calculation of this quantity for this process. It agrees with the general expression \cite{deFlorian:2001zd} relating $B^{(2)}$ to a one loop amplitude.

As well as being of phenomenological interest in their own right (numerical analyses have already been presented \cite{Harlander:2010cz}), our results can form part of a differential \nnlo{} calculation, perhaps along the lines of Ref.~\cite{Catani:2009sm}.

\paragraph{Acknowledgments}
I am grateful to Robert Harlander for many productive discussions throughout the project, and also to Massimiliano Grazzini for useful comments on the manuscript. This research was supported by the US Department of Energy under contract DE-FG03-91ER40662.

\end{document}